\documentclass[prc,aps,superscriptaddress]{revtex4}
\draft
\usepackage{graphicx}
\usepackage{color}
\usepackage{mathtools}
\begin{document}

\title{Single-nucleon properties and pair correlations in nuclear matter from chiral two- and three-nucleon forces}       
\author{       
Francesca Sammarruca}                                                           
\affiliation{ Physics Department, University of Idaho, Moscow, ID 83844-0903, U.S.A. 
}
\author { Herbert M{\" u}ther      }                                                  
\affiliation{ Institut f{\" u}r Theoretische Physik, Universit{\"a}t T{\" u}ubingen, D-72076 T{\" u}bingen, Germany
}
\author{       
Ruprecht Machleidt}                                                           
\affiliation{ Physics Department, University of Idaho, Moscow, ID 83844-0903, U.S.A. 
}
\date{\today} 
\begin{abstract}
We investigate single-particle properties in infinite nuclear matter using a variety of interactions. One of the focal points is to study the impact of chiral three-nucleon forces on the nucleon self-energy and related quantities, such as spectral function and momentum distribution. We also present results for pairing correlations in nuclear matter.
We find characteristic and systematic differences between the predictions obtained with the (softer) chiral interactions and those based on one-boson-exchange or phenomenology. 

\end{abstract}
\maketitle 
        
\section{Introduction} 
\label{Intro} 

 Investigations of single-particle properties in nuclear matter and nuclei, such as spectral functions and momentum distributions, are important and insightful. Studying nucleon dynamics inside nuclei provides, for instance, information about the degree of nucleon correlations, which play a crucial role in nuclear systems. In fact, the presence of correlations is the reason why the mean-field picture of a nucleus is, often times, inadequate. 

A ``signature" of nucleon interaction in the nuclear medium is the formation of quasi-bound two-nucleon states and, thus, the presence of pairing correlations. Most commonly, pairing effects in nuclear physics are understood as referring to the proton-proton ($pp$) or neutron-neutron ($nn$) pairing observed in even-even nuclei. In nuclear matter -- the system that we will use to probe these effects -- there exist non-trivial solutions of the BCS equation in the ($T=1$)  $^1S_0$ partial wave, especially at low density. However, no clear evidence for pairing effects in the $pn$ channel has been found, which may appear surprising because the attraction in the nucleon-nucleon interaction is much stronger in the $T=0$ $S$-wave -- hence the existence of the deuteron. Reasons for the suppression of $pn$ correlations in nuclei have been discussed in Ref.~\cite{MP19}.

The above issues are fundamentally important for understanding the dynamics of nucleons in nuclei. Although an idealized system, homogeneous nuclear matter is a convenient ``test bench" from which to make a transition to finite nuclei~\cite{MP19}. The single-particle properties we examine here are not directly observable while being notoriously interaction dependent.
Therefore, it is crucial to perform careful analyses of model dependence, including state-of-the-art few-nucleon interactions based on chiral effective field theory (EFT). The latter is by now established as the most favorable theory to describe low-energy nuclear physics while maintaing a strong link with the true theory of strong interactions, QCD, through consistency with the symmetry of low-energy QCD.

The study of momentum distributions in nuclear matter and nuclei, particularly for the purpose of discussing short-range correlations (SRC), has a long history. References~\cite{FGM02,SP86,BCdALS86,CdAPS91,Ciof,MP99,MP00,A+13,src2015,A+16,MSVM19} attempt to provide a broad, although non-exhaustive list. 

Our interest in correlations includes both SRC and pairing correlations. 
In this paper, we will address pairing gaps in various channels. Moreover, we will investigate how the strength of correlations and the range of energy or momenta over which they are significant vary with the interaction~\cite{MD05}. The impact of chiral three-nucleon forces (3NFs) is another issue we will address.

In summary, the novel aspects of this work include:
\begin{enumerate}
\item Utilizing high-quality chiral interactions.
\item Comparing predictions, from self-energies to pairing properties, with those obtained from a traditional meson-theoretic nucleon-nucleon ($NN$) potential. Traditional potentials are still in use, such that this comparison is appropriate and interesting.
\item Addressing order-by-order patterns of the predictions, as appropriate for the chiral EFT framework.
\item Studying the impact of chiral 3NFs. To the best of our knowledge, these aspects have not yet been inverstigated.

\end{enumerate}

First, we describe the technical aspects of our approach, starting from the input two-body forces (2NFs) and 3NFs, and proceeding to the nucleon self-energy, spectral function, momentum distribution, and gap function in nuclear matter. The self-energy obtained from the lowest order of the hole-line expansion, namely the Brueckner-Hartree-Fock (BHF) approximation, is complemented with the hole-hole contribution calculated perturbatively~\cite{FGM02}.

Note that momentum densities are not observables.
Recent claims that momentum distributions in nuclei, with particularly emphasis on SRC, can be measured have stimulated considerable interest in the subject.  Although not new, this discussion has resurfaced in conjunction with inclusive or exclusive high-momentum transfer electron scattering experiments.~\cite{FS88,Frank93,Tang,CLAS,CLAS2,CLAS3,Pia+,Eg+06,Shneor,Subedi,Baghda,Pia13,Korover,Hen+17,CT+,Atkwim19}
We will take this opportunity to include  a brief discussion of the issue.

A summary, our conclusions, and future plans are found in Sect.~\ref{Concl}.

\section{Description of the formalism} 
\label{calc}

In this section, we provide details of our calculations, starting with the input two- and three-nucleon chiral forces and proceeding to the self-energy in nuclear matter, including contributions from hole-hole ladders.

 \subsection{The two-nucleon forces}  
\label{II} 

The $NN$ potentials employed in this work are part of a set which spans five orders in the chiral EFT expansion, from leading order (LO) to fifth order (N$^4$LO)~\cite{EMN17}. For the construction of these potentials, the same power counting scheme and regularization procedures are applied through all orders, making this set of interactions more consistent than previous ones.  Another novel and important aspect in the construction of these new potentials is the fact that the long-range part of the interaction is fixed by the $\pi N$ low-energy constants (LECs) as determined in the recent and very accurate analysis of Ref.~\cite{Hofe+}. In fact, for all practical purposes, errors in the $\pi N$ LECs are no longer an issue with regard to uncertainty quantification. Furthemore, at the fifth (and highest) order, the $NN$ data below pion production threshold are reproduced with excellent precision ($\chi ^2$/datum = 1.15).

Iteration of the potential in the Lippmann-Schwinger equation requires cutting off high-momentum components, consistent with the fact that chiral perturbation theory amounts to building a low-momentum expansion. This is accomplished through the application of a regulator function for which the non-local form is chosen:
\begin{equation}
f(p',p) = \exp[-(p'/\Lambda)^{2n} - (p/\Lambda)^{2n}] \,,
\label{reg}
\end{equation}
where $p' \equiv |{\vec p}\,'|$ and $p \equiv |\vec p \, |$ denote the final and initial nucleon momenta in the two-nucleon center-of-mass system, respectively. In the present applications, we will 
consider values for the  cutoff parameter $\Lambda$ between 450 and 500 MeV. 
The potentials are relatively soft as confirmed by the 
Weinberg eigenvalue analysis of Ref.~\cite{Hop17} and in the context of the perturbative calculations of infinite matter of  Ref.~\cite{DHS19}.

\subsection{The three-nucleon forces} 
\label{III}

Three-nucleon forces first appear at the third order of the chiral expansion (N$^2$LO). At this order, the 3NF consists of three contributions~\cite{Epe02}: the long-range two-pion-exchange (2PE) graph, the medium-range one-pion exchange (1PE) diagram, and a short-range contact term. 

For nuclear matter calculations, these 3NFs can be expressed as density-dependent effective two-nucleon interactions as derived in Refs.~\cite{holt09,holt10}. They are represented in  terms of the well-known non-relativistic two-body nuclear force operators and, therefore, can be conveniently incorporated in the usual $NN$ partial wave formalism and the particle-particle ladder approximation for computing the EoS.       
         
The effective density-dependent two-nucleon interactions at N$^2$LO consist of six one-loop topologies. Three of them are generated from the 2PE graph of the chiral 3NF and depend on the LECs $c_{1,3,4}$, which are already present in the 2PE part of the $NN$ interaction. Two one-loop diagrams are generated from the 1PE diagram, and depend on the low-energy constant $c_D$. Finally, there is the one-loop diagram that involves the 3NF contact term, with LEC $c_E$. 

The complete 3NF beyond N$^2$LO is very complex and was neglected in nuclear structure studies of the past.
 However, in recent years, the 3NF at N$^3$LO has been derived~\cite{Ber08} and applied in some nuclear many-body systems~\cite{Tew13,Dri16,DHS19,Heb15a}. 
The contributions to the subleading chiral 3NF include: the 2PE topology, which is the longest-range component of the subleading 3NF, the two-pion-one-pion exchange topology, and the ring topology, generated by a circulating pion which is absorbed and re-emitted from each of the three nucleons. 

Direct inclusion of the subleading chiral 3NF is very challenging for many-body calculations. However, similar to the leading 3NF, the contributions of the 3NF at N$^3$LO can be conveniently expressed in the form of density-dependent effective two-nucleon interactions, as derived in Refs.~\cite{Kais18,Kais19}. Those contributions will be included in future work.

\subsection{Solving the Bethe-Goldstone equation for arbitrary starting energy}

The Bethe-Goldstone (BG) equation for two particles in an uncoupled state with total center-of-mass momentum $\vec{P}$ and initial relative momentum $\vec{q_0}$ in nuclear matter with Fermi momentum $k_F$ is, after angle-averaging of the 
Pauli operator, $Q$:
\begin{equation}
G_{P,\omega,k_F}(q,q_0) = V(q,q_0) + \int_0^{\infty} dk k^2 \frac{V(q,k)Q(k,P,k_F)G(k,q_0)}
{z - (\epsilon(\vec{P} + \vec{k}) + \epsilon(\vec{P} - \vec{k}) ) + i\delta }  \; ,
\label{BG}
\end{equation}
to be solved for each angular momentum state. 
The two particles have initial momenta in the nuclear matter rest frame equal to $\vec{p_1}=\vec{P} + \vec{q_0} $ and 
 $\vec{p_2}=\vec{P} - \vec{q_0} $, respectively. The starting energy $z$ in Eq.~(\ref{BG}) is given by 

\begin{equation}
 z= \omega +  \epsilon(\vec{P} - \vec{q_0})  \; ,
\label{starte}
\end{equation}

where $-\infty \le \omega \le \infty $ and $\epsilon$ is the single-particle energy defined, for a generic momentum $\vec{K}$, as 

\begin{equation}
  \epsilon(\vec{K})= \frac{K^2}{2m} + U(\vec{K})  = \frac{K^2}{2m^*} + U_0 \; ,
\label{spe}
\end{equation}
where $U$ is the single-particle potential and the expression on the right is the parameterized form 
in terms of (pre-determined) self-consistent parameters $m^*$ and $U_0$. Thus, 

\begin{equation}
  z = \omega +\frac{p_2^2}{2m^*} + U_0 \; .
\label{starte2}
\end{equation}

Using Eq.~(\ref{spe}) and Eq.~(\ref{starte2}), the denominator of Eq.~(\ref{BG}) can be written as 
\begin{equation}
z - (\epsilon(\vec{P} + \vec{k}) + \epsilon(\vec{P} - \vec{k}) ) = 
\frac{1}{m^*} \Big ( m^*(\omega - U_0) + \frac{p_2^2}{2} -P^2 -k^2  \Big ) \; ,
\label{den}
\end{equation} 
from which it is easy to see that the singularity occurs at 
\begin{equation}
q_s^2 = 
 m^*(\omega - U_0) + \frac{p_2^2}{2} -P^2 \; . 
\label{den}
\end{equation} 
We then proceed to solve the equation 
\begin{equation}
G_{P,\omega,k_F}(q,q_0) = V(q,q_0) + m^*\int_0^{\infty} dk k^2 \frac{V(q,k)Q(k,P,k_F)G(k,q_0)}
{q_s^2 - k^2 + i\delta }  
\label{BG2}
\end{equation}
with standard subtraction and matrix inversion techniques to obtain $G(q,q_0)$,  from which we can easily extract $G(q_0,q_0)$, and finally the (BHF) nucleon self-energy:
\begin{equation}
\Sigma(p_1,\omega) = \int d^3p_2 G(\vec{q_0}(\vec{p_1},\vec{p_2}),\vec{q_0}(\vec{p_1},\vec{p_2})) \;.
\label{sigma}
\end{equation} 

Following Ref.~\cite{FGM02}, the hole-hole contribution can be written as:

\begin{equation}
\Delta\Sigma_{2h1p}(p_1,\omega) =  \int_{k_F}^{\infty} d^3p \int_0^{k_F}d^3h_1 d^3h_2   
\frac{<p_1,p|G|h_1 h_2>^2}
{\omega + \epsilon_p - \epsilon_{h_1} - \epsilon_{h_2}  + i\delta }  \; .
\label{dsig}
\end{equation}
We recall that an imaginary part of the BHF self-energy being different from zero signifies particle-particle excitations, whereas the imaginary part of $\Delta\Sigma$  plays the same role for hole-hole excitations.
Actually, we use Eq.~(\ref{dsig}) to obtain the imaginary part of $\Delta\Sigma_{2h1p}$, and derive the real part from a dispersion relation~\cite{FGM02}.

The total self-energy is then 
\begin{equation}
\Sigma_{tot}(p_1,\omega) = \Sigma(p_1,\omega) +\Delta\Sigma_{2h1p}(p_1,\omega) \; ,
\label{sig_tot}
\end{equation}
from which 
the spectral functions $S_{h(p)}$ for hole (particle) strength is readily obtained:
\begin{equation}
S_{h(p)}(p_1,\omega) = +(-) \frac{1}{\pi} \frac{\mbox{Im} \Sigma_{tot}(p_1,\omega)}{\Big (\omega - p_1^2/2m - \mbox{Re} \Sigma_{tot}(p_1,\omega) \Big )^2 + \Big (\mbox{Im} \Sigma_{tot}(p_1,\omega) \Big )^2 } \; ,
\label{spctr}
\end{equation}
for $\omega < \epsilon_F$ ($\omega > \epsilon_F$).

The occupation probability or momentum density is then calculated integrating the spectral distribution of the hole states:
\begin{equation}
n(p_1) = \int_{-\infty}^{\epsilon_F} d\omega S_{h}(p_1,\omega) \;,
\label{momdis}
\end{equation}
where $\epsilon_F$ is the Fermi energy.

Note that, in the mean field approximation, the spectral functions reduce to the well-known expressions:
\begin{equation}
S_p(p_1,\omega) = \Theta(p_1-k_F) \delta(\omega - \epsilon(p_1)) \; , 
\end{equation} 
and
\begin{equation}
S_h(p_1,\omega) = \Theta(k_F-p_1) \delta(\omega - \epsilon(p_1)) \; , 
\end{equation} 
where 
\begin{equation}
  \epsilon(\vec{p_1})= \frac{p_1^2}{2m} + \mbox{Re}\Sigma(p_1,\omega) 
\label{upot}
\end{equation}
with the real part of the self-energy calculated on shell.

\subsection{Pairing Gap}
\label{sec.pairing}
In the last subsection, we have discussed -- separately --  the particle-particle ($pp$) and hole-hole ($hh$) contributions to the self-energy and spectral functions of nucleons. While the former contribution is contained in the solution of the Bethe-Goldstone equation and the BHF part of the self-energy $\Sigma(p,\omega)$, the $hh$ contribution has been evaluated in a perturbative way leading to $\Delta\Sigma_{2h1p}(p,\omega)$. In the framework of the self-consistent Green's functions (SCGF) approach~\cite{Dickhoff}, both contributions should be treated in a non-perturbative way. This implies that instead of solving the Bethe-Goldstone equation, Eq.~(\ref{BG}), one has to solve the generalized $T$-matrix equation including $pp$ and $hh$ contributions at all orders. It turned out, however, that conventional methods to solve this ladder equation for infinite nuclear matter or finite nuclei using realistic $NN$ interactions are plagued by the presence of pairing effects or quasi-bound states in various partial waves (see e.g. Ref.~\cite{ramos1}). Techniques have been developed to avoid this difficulty by investigating nuclear matter at finite temperature and inspecting the $T \rightarrow 0$ limit only after a self-consistent solution is obtained~\cite{frick05}.

The occurence of quasi-bound states or pairing phenomena can be very nicely isolated by evaluating the generalized $T$ matrix or the two-particle Green's function    
in terms of discrete eigenvalues and eigenfunctions of a two-particle Hamiltonian~\cite{rubts17,NM1}. 
This includes both $pp$ and $hh$ states and corresponds to the Hamiltonian of the $pphh$ Random Phase Approximation (RPA):
\begin{equation}
\left(\begin{array}{cc} H^0_{p} + V_{pp} & V_{ph} \\ -V_{hp} & H^0_h-V_{hh}\\ \end{array}\right) \,. \label{eq:pphhrpa}
\end{equation}
In this matrix $H^0_p$ represents the single-particle part of the $pp$ Hamiltonian, which means that it can be written as
$$
H^0_p = \sum_{p1,p2} \left(\tilde\varepsilon_{p1} + \tilde\varepsilon_{p2}\right) \vert p_1p_2 \rangle\langle p_1p_2\vert\,,
$$
with the single-particle energies
$$
\tilde\varepsilon_p = \varepsilon_p - \varepsilon_F
$$
rescaled by the Fermi energy $\varepsilon_F$. The corresponding definition applies to the $hh$ single-particle Hamiltonian $H^0_h$, while the two-body part of the RPA Hamiltonian is defined in terms of matrix elements of the form
$$
V_{ph} \Longleftrightarrow \langle p_1p_2\vert V \vert h_1h_2\rangle\,,
$$
and analogous expressions for $V_{pp}$, $ V_{hp}$ and $V_{hh}$. If, for the moment, we restrict the discussion to $pp$ and $hh$ states of nuclear matter with center of mass momentum $P=0$, the $pp$ states are identified as two-particle states with momenta $\vec k$ and $-\vec k$ with $\vert \vec k\vert > k_F$. This means that, after a partial wave decomposition, such a state is characterized by one wavenumber $k$ larger than the Fermi momentum $k_F$ for the $pp$ states and smaller than $k_F$ for the $hh$ states. 

Rubtsova {\it et al.} demonstrated~\cite{rubts17} that, after discretizing the momentum variable $k$, a nontrivial solution of the BCS equation for nuclear matter in a specific partial wave is signaled by a pair of complex conjugate eigenvalues for the corresponding $pphh$ RPA Hamiltonian. Since the Hamiltonian in Eq.~(\ref{eq:pphhrpa}) is non-hermitian but real, it may have pairs of complex  eigenvalues, $E_\beta$ and $E_\beta^*$, with conjugated eigenfunctions $\vert \Phi_\beta \rangle$ and $\vert \Phi_\beta^*\rangle$.  In fact, the imaginary part of these eigenvalues is the pairing gap at the Fermi surface:
\begin{equation}
\vert \Im E_\beta \vert = \Delta (k_F)\,. \label{eq:imagval} 
\end{equation}
Furthermore it has been observed that the wavefunction of the bound state is proportional to the function of the pairing gap,
$$
\left| \langle k \vert \Phi_\beta \rangle\right| \sim \Delta(k)\,,
$$
which means that the bound-state of the $pphh$ RPA is rather close to the solution of the BCS gap equation. The latter can be written as
\begin{equation}
\left(\begin{array}{cc} \tilde H^0_{p} + V_{pp} & V_{ph} \\ V_{hp} & \tilde H^0_h + V_{hh}\\ \end{array}\right)\left(\begin{array}{c} \langle p \vert \chi \rangle \\ \langle h \vert\chi \rangle \\ \end{array}\right) = 0 \; ,  \label{eq:pphhbcs1}
\end{equation}
with
\begin{equation}
\tilde H^0_p = \sum_{k>k_F} 2\sqrt{\left(\varepsilon_k -\varepsilon_F\right)^2 + \Delta(k)^2}\vert k\rangle\langle k\vert \,, \label{eq:pphhbcs2}
\end{equation}
and a corresponding definition for the $hh$ part $\tilde H^0_h$. The solution of the homogeneous Eq.~(\ref{eq:pphhbcs1}),  $\chi$, is to be interpreted as
\begin{equation}
\left| \langle k \vert \chi \rangle\right| = \frac{\Delta(k)}{2\sqrt{\left(\varepsilon_k -\varepsilon_F\right)^2 + \Delta(k)^2}}\,.\label{eq:pphhbcs3}
\end{equation} 
In fact, the representation of the BCS equation for nuclear matter in Eqs.~(\ref{eq:pphhbcs1} - \ref{eq:pphhbcs3}) leads to a very efficient solution of the non-linear BCS equation: Assume that $\Delta = 0$ and determine the eigenvalues of the matrix in Eq.~(\ref{eq:pphhbcs1}). A nontrivial solution for the gap function is only obtained, if the lowest eigenvalue is below zero energy (compare discussion of Eq.~(\ref{eq:imagval})). The complete function $\Delta(k)$ can be obtained from an iterative solution of Eqs.~(\ref{eq:pphhbcs1} - \ref{eq:pphhbcs3}) until the lowest eigenvalue occurs at zero energy. This formulation of the gap equation in terms of a bound state, and the gap function needed to shift this bound state to zero energy is close to the discussion of pairing instabilities in  Ref.~\cite{Dickhoff}.

\section{Results}
\label{res} 

\begin{figure*}[!t] 
\centering
\hspace*{-1.5cm}
\includegraphics[width=7.5cm]{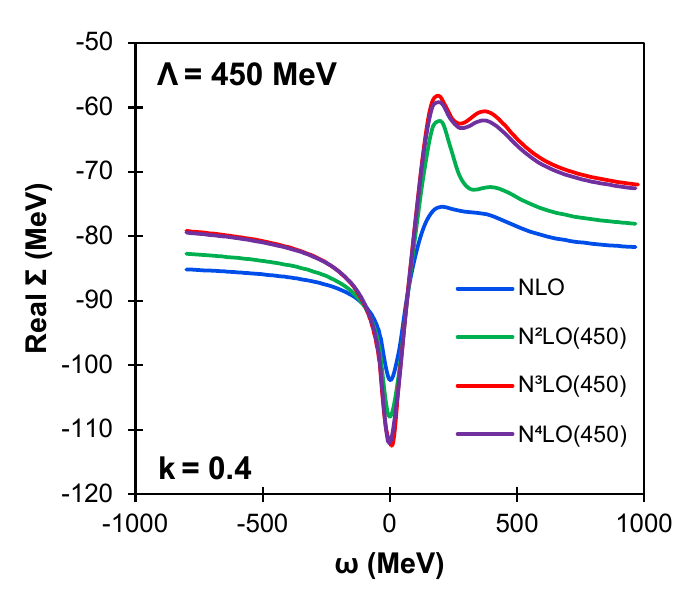}\hspace{0.01in} 
 \includegraphics[width=7.5cm]{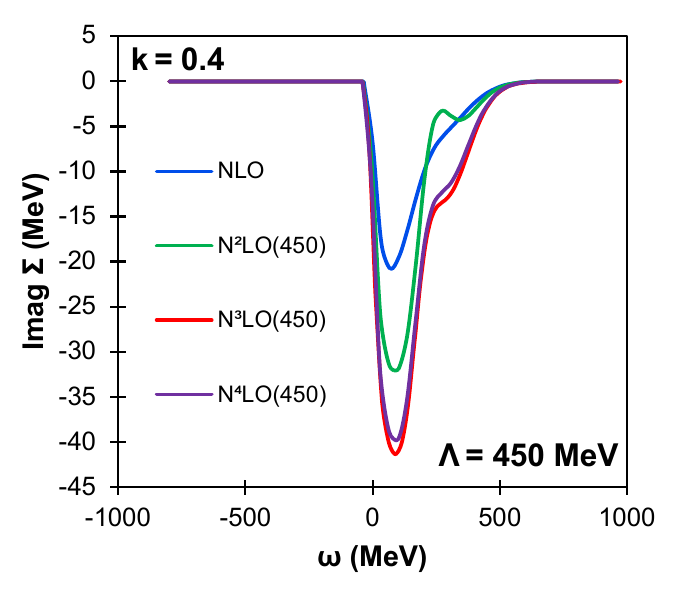}\hspace{0.01in }
\vspace*{0.05cm}
 \caption{(Color online) The real and the imaginary parts of $\Sigma$ {\it vs.} the energy and for fixed momentum, given in units of the Fermi momentum. The predictions are obtained with the chiral potentials of Ref.~\cite{EMN17} from second to fifth order.  } 
 \label{sig_obo}
\end{figure*}

We begin with discussing  the real and imaginary parts of the (BHF) self-energy, $\Sigma(k,\omega)$, see  Eq.~(\ref{sigma}). We will show predictions from second (NLO) to fifth (N$^4$LO) order of the chiral expansion~\cite{EMN17}, see Fig.~\ref{sig_obo}. Note that,  from this point forward, $k$ is a single-particle momentum.
Unless otherwise specified, the regulator $\Lambda$ is taken to be 450 MeV, and the Fermi momentum is equal to 1.35 fm,$^{-1}$  corresponing to a density of 0.166 fm,$^{-3}$ close to the empirical saturation density. 

Figure~\ref{sig_obo} shows the real and imaginary parts of $\Sigma$ as functions of the energy and for fixed momentum, which is given in units of the Fermi momentum. The predictions are obtained with 2NFs  only. Under these conditions, we see good convergence pattern towards the fifth order. We observed this to be true irrespective of the particular momentum.

Inspecting the imaginary part of $\Sigma(k,\omega)$ in the right panel of Fig.~\ref{sig_obo}, one finds that the absolute value of the imaginary part, which is a measure for the importance of correlation effects, increases with the order of the chiral expansion. The imaginary part evaluated at N$^2$LO shows a second minimum, a structure which persists but is damped at the higher orders of the chiral expansion.

The energy-dependent contribution to the real part of the self-energy can be evaluated from the imaginary part by a dispersion relation, and the limit 
\begin{equation}
\lim_{\omega \to -\infty} \Sigma(k,\omega) = \Sigma_{HF}(k)\,,\label{eq:hflimit}
\end{equation}
corresponds to the contribution of the bare $NN$ interaction to the self-energy, namely the Hartree-Fock approximation for the self-energy. Because all interaction models were fitted to the $NN$ phase shifts, the real parts for the self-energy are close to each other at energies around the Fermi energy ($\omega \approx \varepsilon_F$). This implies that the Hartree-Fock (HF) limit at N$^4$LO is less attractive than the one at NLO. There is more attraction due to $pp$ correlations in the self-energy calculated at N$^4$LO than in the one evaluated at NLO (see also the discussion of the imaginary parts of $\Sigma (k,\omega)$ above).

\begin{figure*}[!t] 
\centering
\hspace*{-1.5cm}
\includegraphics[width=8.0cm]{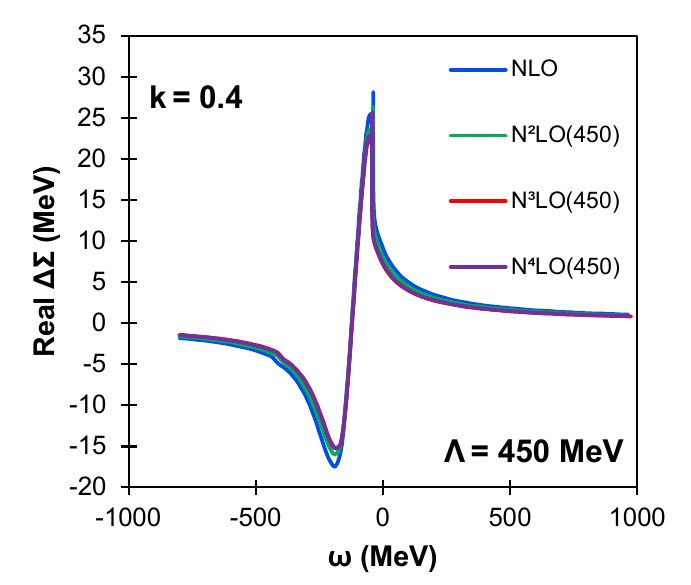}\hspace{0.01in} 
 \includegraphics[width=7.7cm]{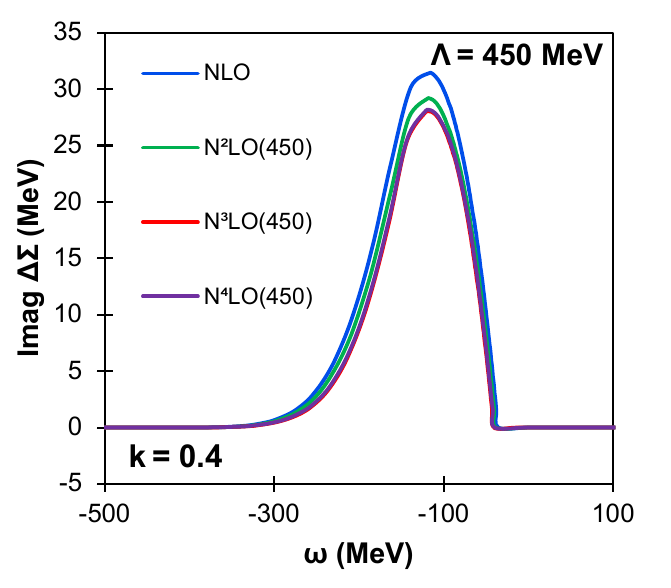}\hspace{0.01in }
\vspace*{0.05cm}
 \caption{(Color online) Same legend as for Fig.~\ref{sig_obo} but for the hole-hole (or $2h1p$) contribution to the total self-energy. } 
 \label{dsig_obo}
\end{figure*}      

A similar description applies to Fig.~\ref{dsig_obo} but for the $hh$ (or $2h1p$) contribution to the total self-energy, $\Delta \Sigma_{2h1p}$,  Eq.~(\ref{dsig}). This $2h1p$ contribution is calculated in terms of the $G$-matrix calculated at a an average starting energy of $hh$ states. The $G$-matrix at these energies is very similar for all the interactions considered, leading to a result that is almost independent of the order of the chiral expansion. 

\begin{figure*}[!t] 
\centering
\hspace*{-1.5cm}
\includegraphics[width=7.7cm]{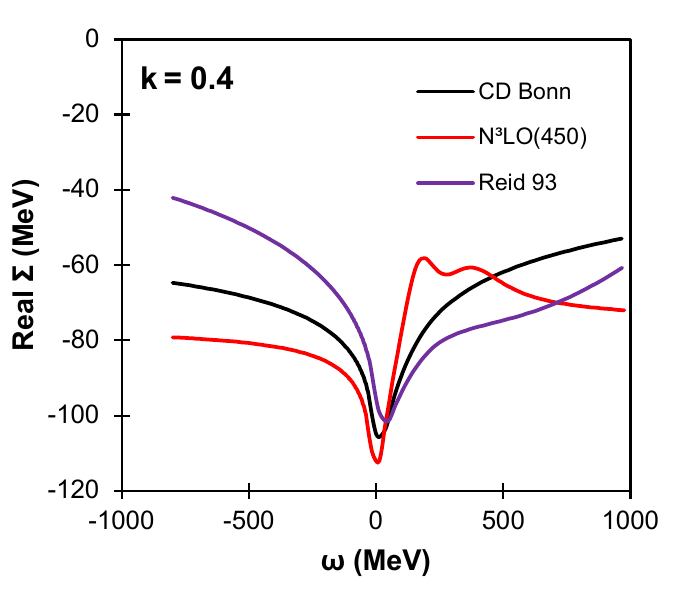}\hspace{0.01in} 
 \includegraphics[width=7.7cm]{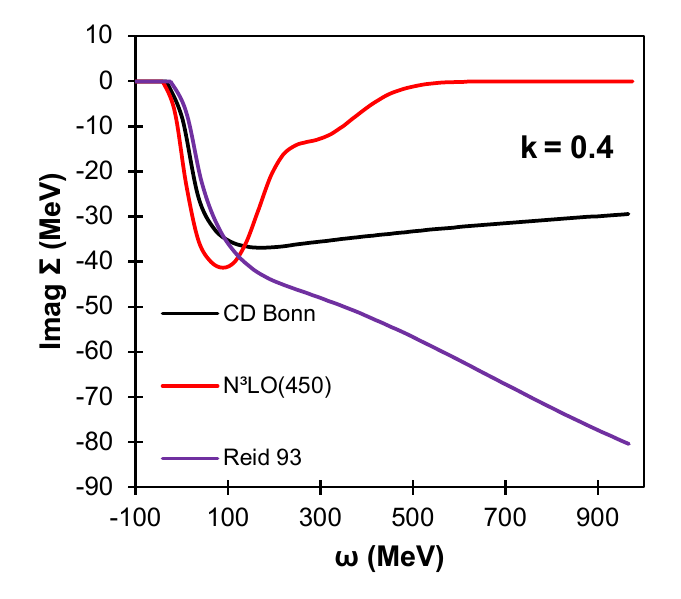}\hspace{0.01in }
\vspace*{0.05cm}
 \caption{(Color online) Comparison beween the predictions for $\Sigma$ obtained with the N$^3$LO(450)  chiral potential, the CD-Bonn $NN$ potential, and the Reid93 local $NN$ potential. The momentum is fixed at 0.4 of the Fermi momentum.} 
 \label{CT1_sig}
\end{figure*}

\begin{figure*}[!t] 
\centering
\hspace*{-1.5cm}
\includegraphics[width=8.0cm]{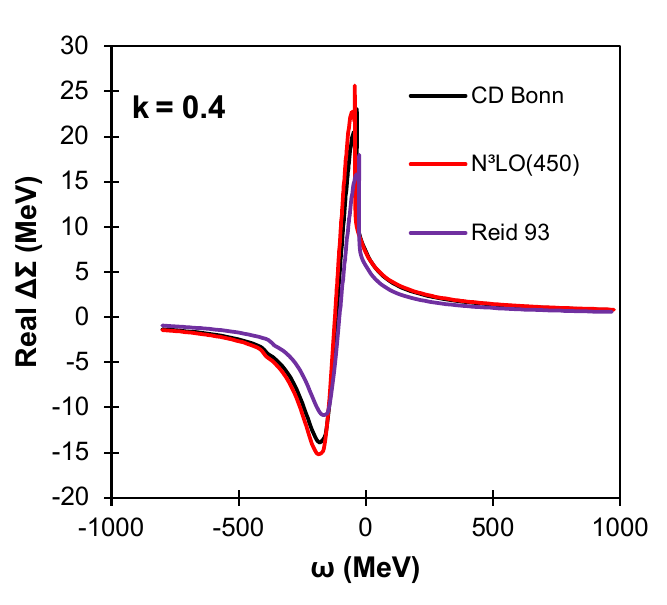}\hspace{0.01in} 
 \includegraphics[width=8.0cm]{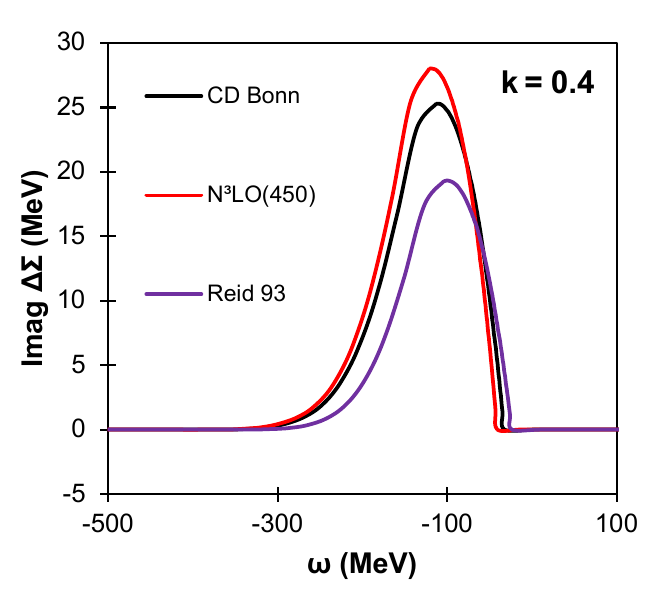}\hspace{0.01in }
\vspace*{0.05cm}
 \caption{(Color online) As in Fig.~\ref{CT1_sig}, but for $\Delta\Sigma$. } 
 \label{CT1_dsig}
\end{figure*}

In Fig.~\ref{CT1_sig} and \ref{CT1_dsig}, predictions for $\Sigma$ and $\Delta\Sigma$ obtained with the N$^3$LO(450), CD-Bonn~\cite{CD}, and the Reid93~\cite{Reid93} $NN$ potentials are compared at fixed momentum. Note that we have chosen Reid93 to represent the generation of high-precision {\it local} potentials, known to be considerably ``harder" than the corresponding generation of non-local potentials. Hugh differences are found especially for the imaginary part of $\Sigma$ at energies above the Fermi energy, as seen in the right panel of Fig.~\ref{CT1_sig}. The absolute value of the imaginary part of $\Sigma$ derived from Reid93 increases even at energies above 1 GeV. This corresponds to a strong energy dependence of the real part of $\Sigma$ presented in the left part of Fig.~\ref{CT1_sig}, which even turns repulsive in the Hartree-Fock limit of $\omega \to -\infty$. Both reflect the strong repulsive core in local potentials, required to describe the phase-shifts in the $S$-waves at energies around and above 300 MeV. The results obtained for the CD-Bonn interaction are in between those from the other two interactions considered in these figures.

The predictions for $\Delta\Sigma$ displayed in Fig.~\ref{CT1_dsig} are much less sensitive to the choice of the interaction (see discussion above). It is worth noting that the results for the imaginary part of $\Delta\Sigma$ are larger for the chiral interaction as compared with those evaluated with CD-Bonn or Reid93. This indicates that the chiral interaction is slightly stronger at smaller momenta, which generate the $2h1p$ contribution to the self-energy.

At this point it may be useful to recall that the large difference between the $pp$ and the $hh$ correlation effects --   reflected in the sizes of the imaginary part resulting from $2p1h$ and $2h1p$ contributions --  motivated classical expansions of the many-body theory, like the hole-line expansion. Also, the $2p1h$ contributions to the self-energy from the (softer) chiral interactions are larger than the $2h1p$ contributions. The difference, however, is much smaller than for traditional potentials such as CD-Bonn or Reid93. This indicates that many-body approaches like the self-consistent Green's function method (SCGF)~\cite{Dickhoff}, which treat particle- and hole-state contributions on the same footing, are preferable in studies of nuclear systems with chiral interactions.

\begin{figure*}[!t] 
\centering
\hspace*{-1.5cm}
\includegraphics[width=8.0cm]{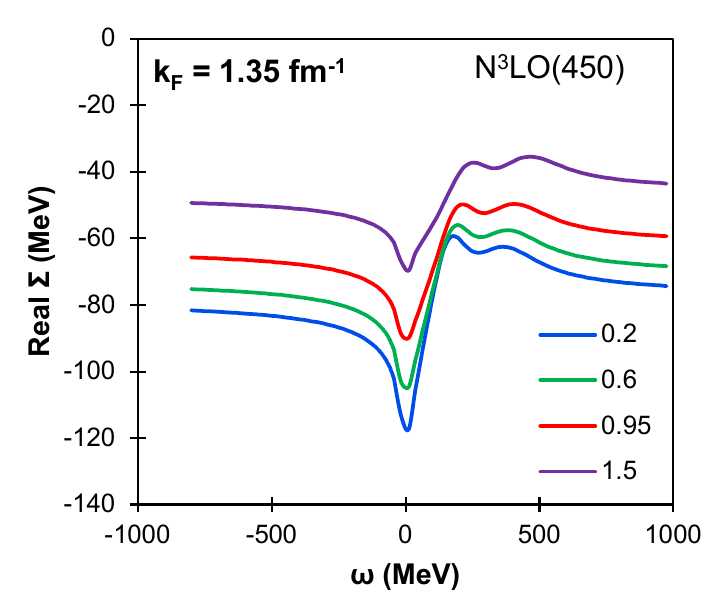}\hspace{0.01in} 
 \includegraphics[width=7.5cm]{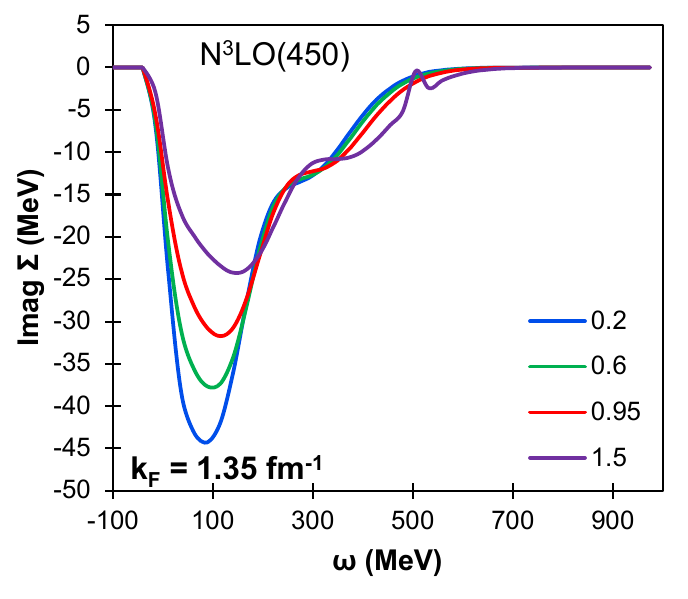}\hspace{0.01in }
\vspace*{0.05cm}
 \caption{(Color online) The momentum dependence of $\Sigma$ and $\Delta\Sigma$, again at fixed density and with the N$^3$LO(450) interaction.  The numbers inside the figure frame are values of the momentum in units of the Fermi momentum.} 
 \label{sig_k}
\end{figure*}

\begin{figure*}[!t] 
\centering
\hspace*{-1.5cm}
\includegraphics[width=8.0cm]{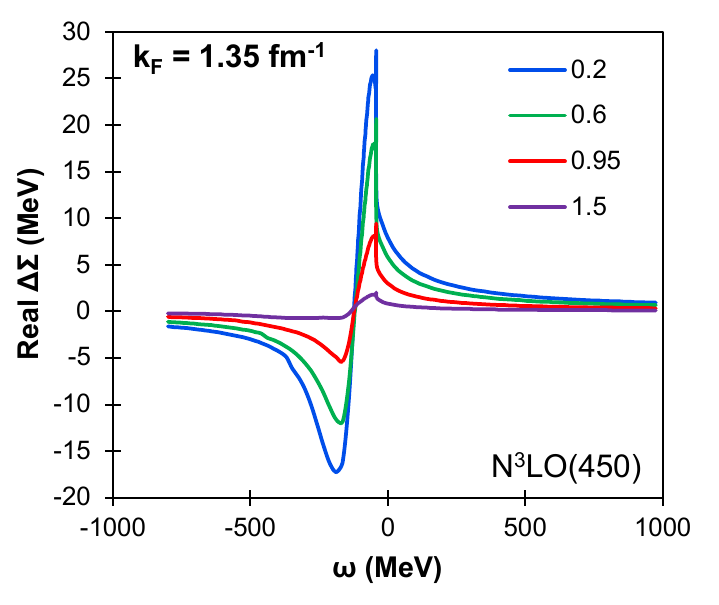}\hspace{0.01in} 
 \includegraphics[width=7.7cm]{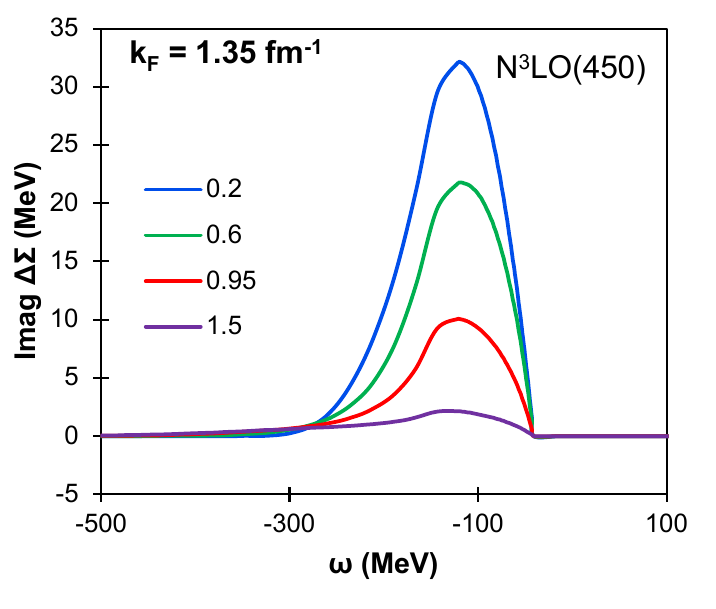}\hspace{0.01in }
\vspace*{0.05cm}
 \caption{(Color online) As in Fig.~\ref{sig_k}, but for $\Delta\Sigma$. }
 \label{dsig_k}
\end{figure*}

The next two figures, Figs.~\ref{sig_k}-\ref{dsig_k}, display the momentum dependence of $\Sigma$ and $\Delta\Sigma$, again at fixed density and with the N$^3$LO(450) interaction, which will be our standard choice unless otherwised specified. For both the real and imaginary parts of $\Sigma$ and $\Delta\Sigma$, the magnitude of the self-energy decreases rapidly with increasing momentum, as to be expected on physical grounds since the self-energy is the ``dressing" acquired by the nucleon as it travels through nuclear matter.

\begin{figure*}[!t] 
\centering
\hspace*{-1.5cm}
\includegraphics[width=7.5cm]{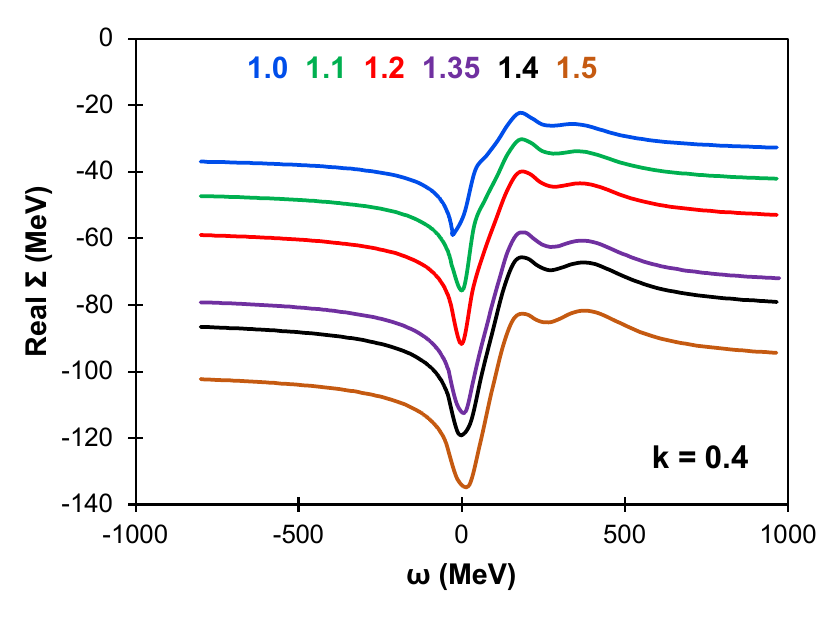}\hspace{0.01in} 
 \includegraphics[width=7.5cm]{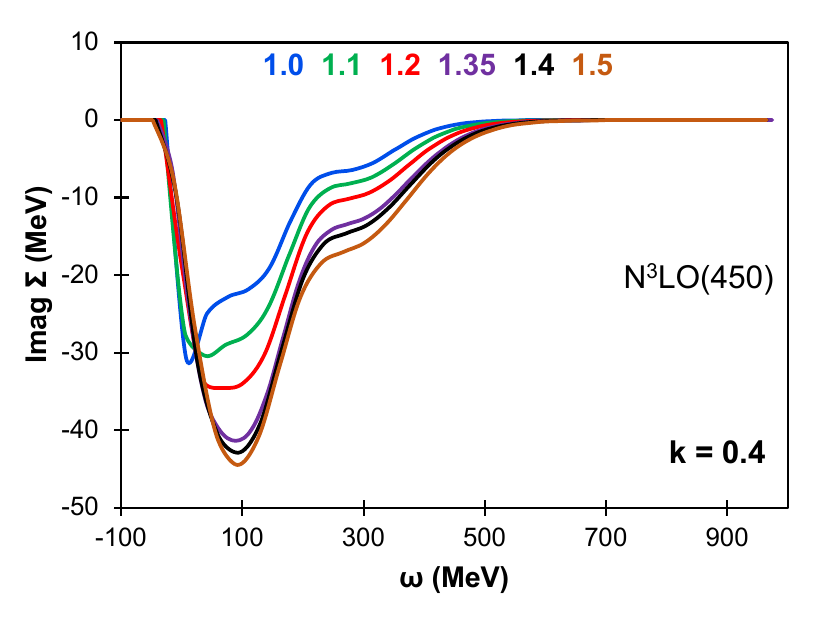}\hspace{0.01in }
\vspace*{0.05cm}
 \caption{(Color online) Real and imaginary parts of  $\Sigma$ for fixed momentum and changing density. The values inside the figure frame are the Fermi momenta in fm$^{-1}$ and correspond to the curve with the same color. } 
 \label{DD}
\end{figure*}

 On the other hand, the self-energy increases in magnitude with increasing density. This is seen in Fig.~\ref{DD}, where 
 we explore the density dependence of the self-energy for fixed momentum. In Fig.~\ref{DD}, the various colors indicate the different Fermi momenta given  inside the figure with colors consistent with the corresponding curves.

\begin{figure*}[!t] 
\centering
\hspace*{-1.5cm}
\includegraphics[width=7.5cm]{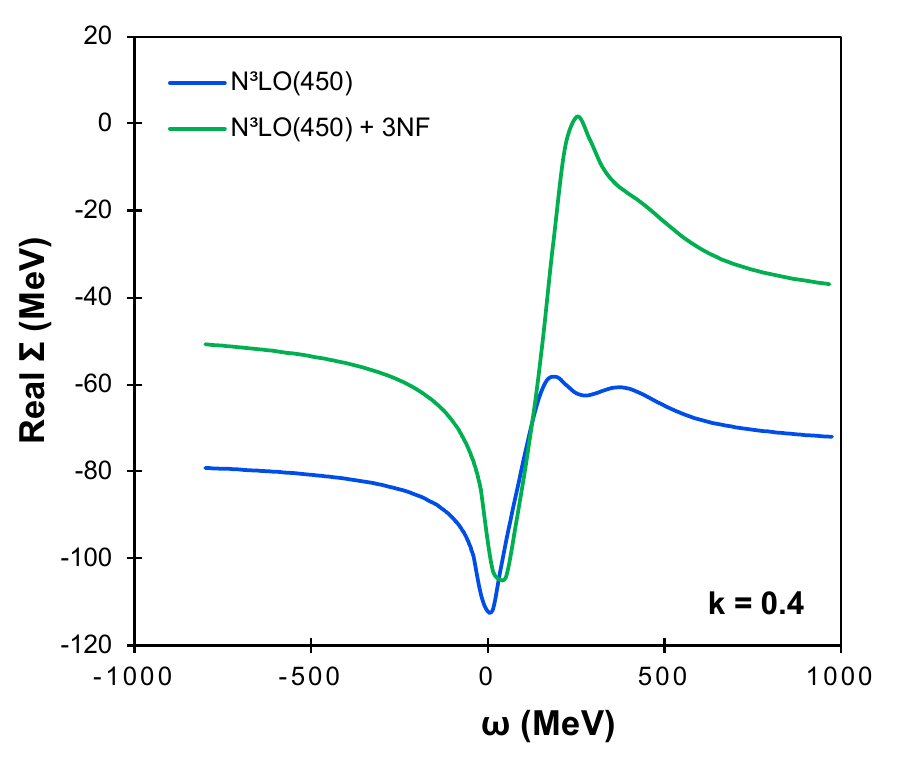}\hspace{0.01in} 
 \includegraphics[width=7.3cm]{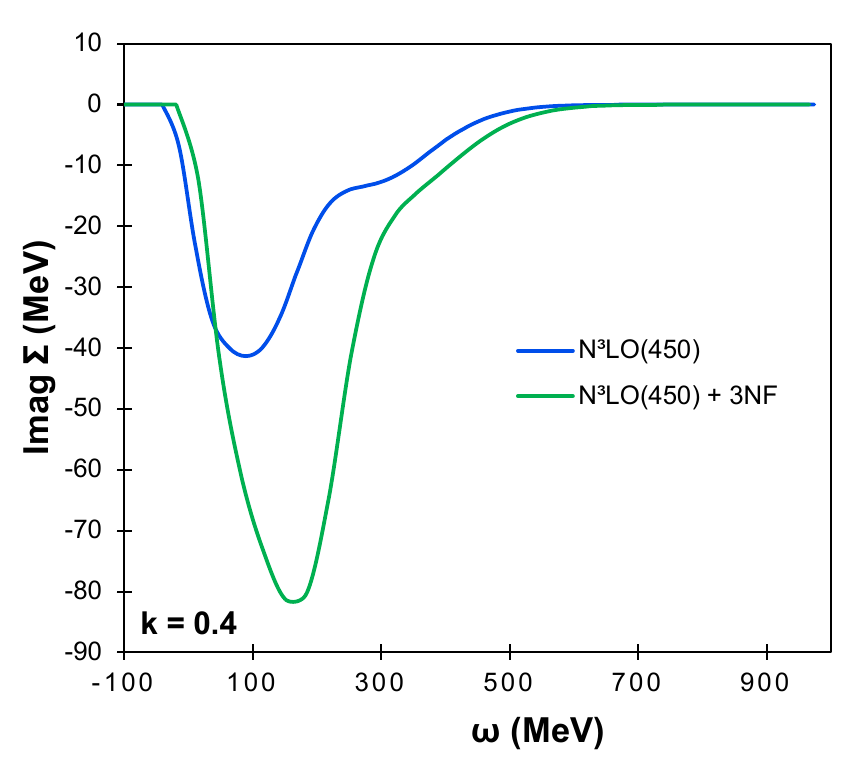}\hspace{0.01in }
\vspace*{0.05cm}
 \caption{(Color online) Real and imaginary parts of the self-energy at fixed momentum, with and without the leading chiral 3NF.  The N$^3$LO(450) potential is used for the 2NF. } 
 \label{sig_3nf_04}
\end{figure*}

\begin{figure*}[!t] 
\centering
\vspace*{0.5cm}
\hspace*{-1.5cm}
\includegraphics[width=7.5cm]{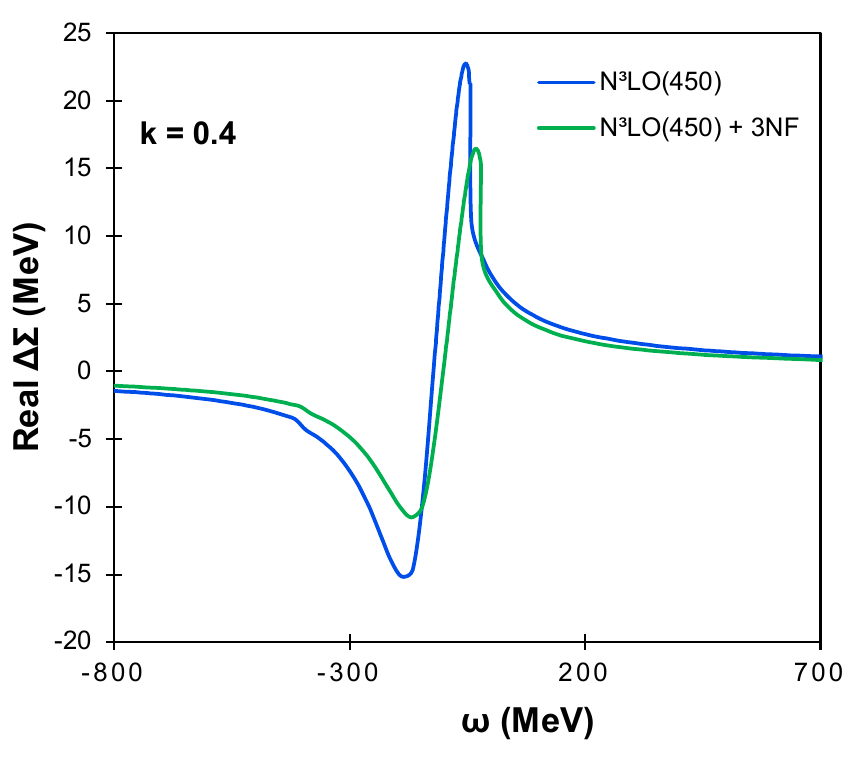}\hspace{0.01in} 
 \includegraphics[width=7.5cm]{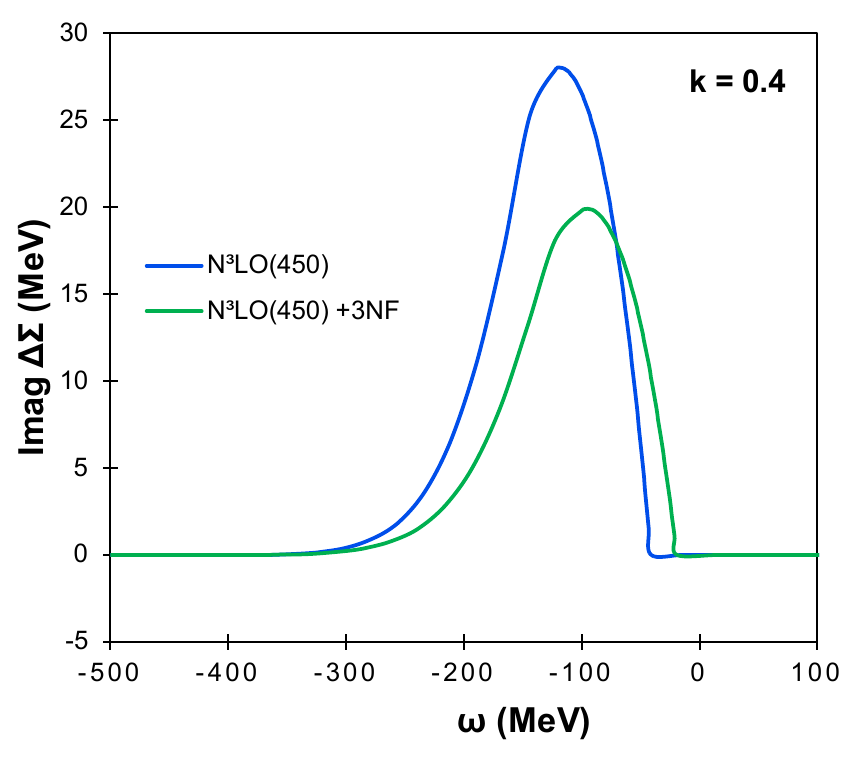}\hspace{0.01in }
\vspace*{0.05cm}
 \caption{(Color online) As in Fig.~\ref{sig_3nf_04} but for $\Delta\Sigma$. } 
 \label{dsig_3nf_04}
\end{figure*}

The next set of results, Figs.~\ref{sig_3nf_04}-\ref{dsig_3nf_04}, addresses the contribution of the leading chiral 3NF, which we include as described in sect.~\ref{III}. 
In Fig.~\ref{sig_3nf_04} we compare predictions for $\Sigma$ at fixed momentum and density with and without 3NF contributions. The impact on $\Sigma$ is very large for both the real and imaginary parts. Note that the repulsive effects of the 3NF reduce the HF contribution to the self-energy (see Eq.~(\ref{eq:hflimit})) by almost a factor 1/2. The contribution to the imaginary part of $\Sigma$ is almost as large as the contributions due to the $NN$ interaction and in the opposite direction compared to the effect on the real part. The contribution of the 3NF  on $\Delta\Sigma$ is substantially smaller and tends to compensate the effects of the 2NF to some extent.

\begin{figure*}[!t] 
\centering
\vspace*{2.0cm}
\hspace*{-1.5cm}
\includegraphics[width=7.5cm]{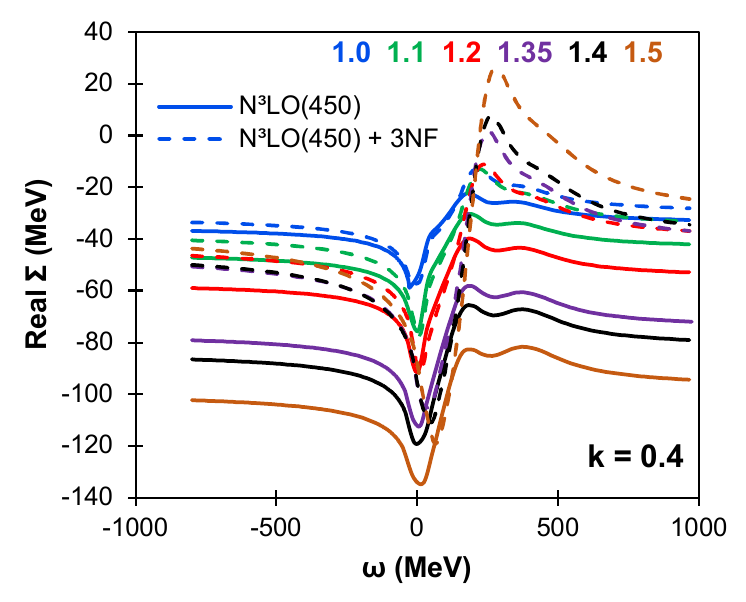}\hspace{0.01in} 
 \includegraphics[width=7.5cm]{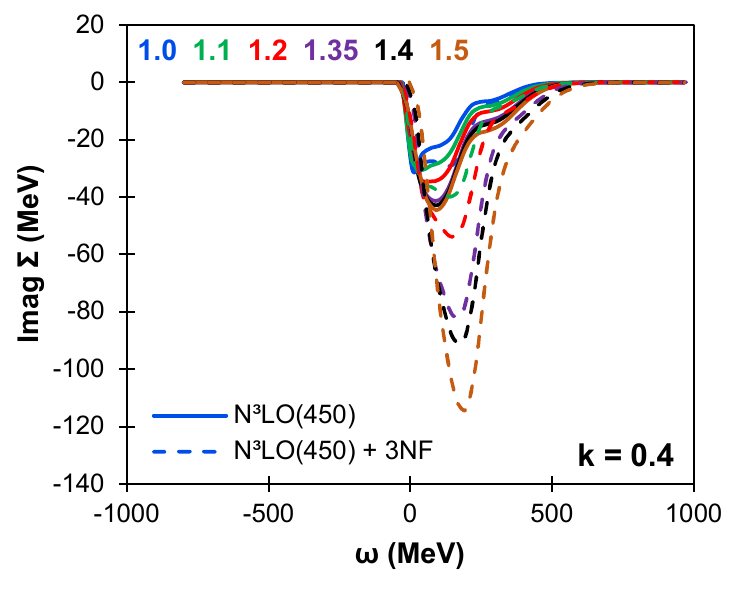}\hspace{0.01in }
\vspace*{0.05cm}
 \caption{(Color online)Real and imaginary parts of  $\Sigma$ for fixed momentum and changing density. The values inside the figure frame are the Fermi momenta in fm$^{-1}$ and correspond to the curve with the same color.  The dashed curves show results including the leading chiral 3NF.}  
 \label{DD_3nf}
\end{figure*}

It is interesting to compare the density dependence of the self-energy with or without 3NF. This comparison is shown in Fig.~\ref{DD_3nf}. While very mild at low density, the impact of the 3NF becomes increasingly strong with increasing density, as to be expected from analyses of 3NF effects in the nuclear matter equation of state at and above saturation.

\begin{figure*}[!t] 
\centering
\vspace*{2.0cm}
\hspace*{-1.5cm}
\includegraphics[width=7.5cm]{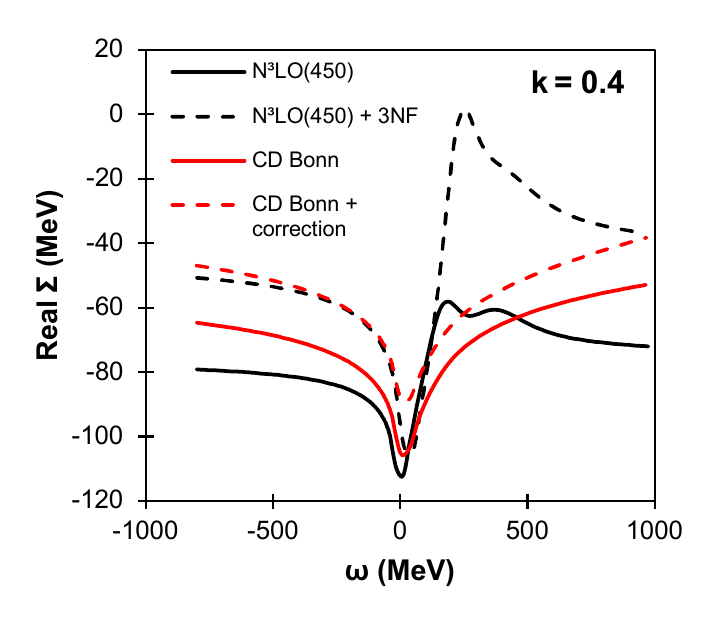}\hspace{0.01in} 
 \includegraphics[width=7.5cm]{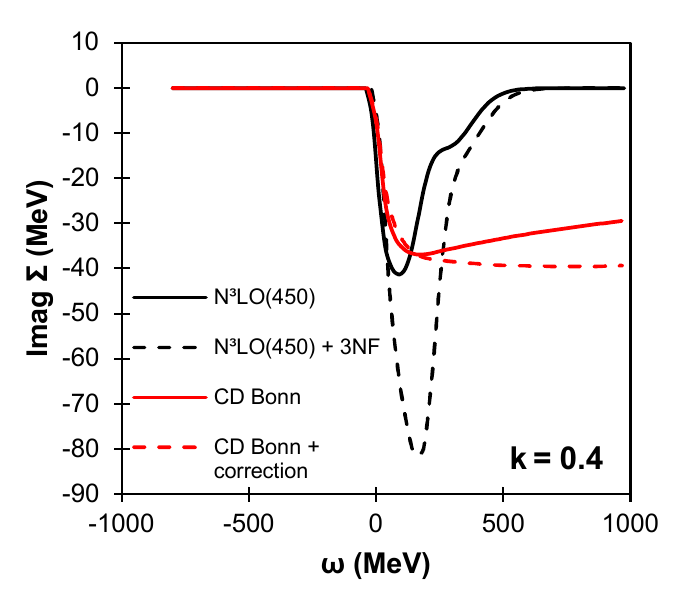}\hspace{0.01in }
\vspace*{0.05cm}
 \caption{(Color online)Real and imaginary parts of  $\Sigma$ for fixed momentum $k=0.4 k_F$ and Fermi momentum $k_F=1.35 $ fm$.^{-1}$ The black dashed (black solid) curve shows results with (without) inclusion of the chiral 3NF. The red dashed (red solid) curves displays the results with (without) relativistic corrections on CD-Bonn. See text for details.}
 \label{CCD_3nf}
\end{figure*}

Three-nucleon forces expressed in terms of a density-dependent 2NF also occur within the meson-exchange model of the $NN$ interaction. As an example, we mention the change of the $NN$ interaction due to the density dependence of the nucleon mass in the in-medium Dirac spinors, as done in the framework of the Dirac-Brueckner-Hartree-Fock approach (DBHF)~\cite{Machxxx}. Interpreting the in-medium spinor as a combination of positive and negative solutions of the Dirac equation, one may argue that the ``Dirac mass" mechanism is approximately equivalent to a 3NF generated by nucleon-antinucleon excitation. To simulate such 3NF, we have calculated the differences between the predictions of a One-Boson-Exchange potential with and without including the density dependence of the Dirac spinors, and then added these differences to the original CD-Bonn predictions. 
Results of this test for the self-energy $\Sigma$ are given in Fig.~\ref{CCD_3nf} and compared to the effect of the leading chiral 3NF on the N$^3$LO(450) predictions. It is interesting to see that the corrections to the real part of the self-energy from the chiral 3NF (on the N$^3$LO(450) predictions) and those from the ``Dirac mass" (on the CD-Bonn predictions) are rather similar. In fact, it has been argued that the Dirac effect may provide a saturation mechanism in nuclear matter when low-momentum interactions are used~\cite{eriklowk}. A similar comparison for the imaginary part of the self-energy, right panel of Fig.~\ref{CCD_3nf}, leads to very different conclusions. While the chiral 3NF leads to an enhancement of the imaginary part at low energies, the Dirac effect added to the CD-Bonn predictions yields larger values at higher energies.

\begin{figure*}[!t] 
\centering
\vspace*{1.0cm}
\hspace*{-1.5cm}
\includegraphics[width=7.0cm]{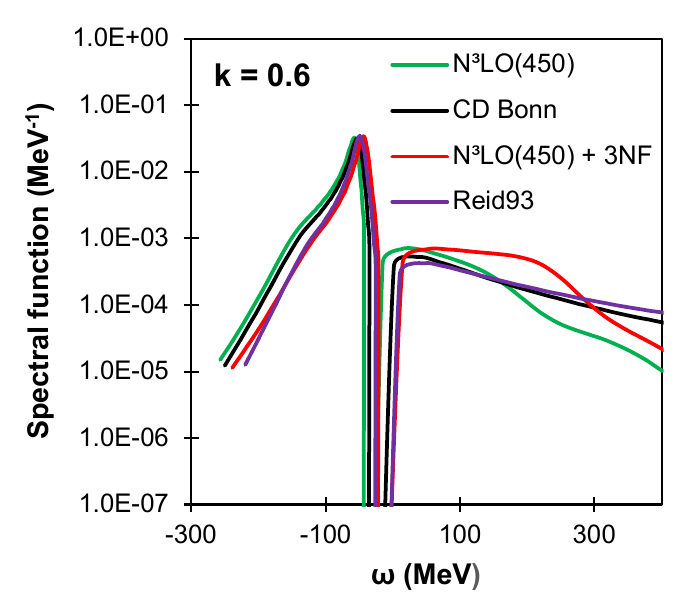}\hspace{0.01in} 
\includegraphics[width=7.0cm]{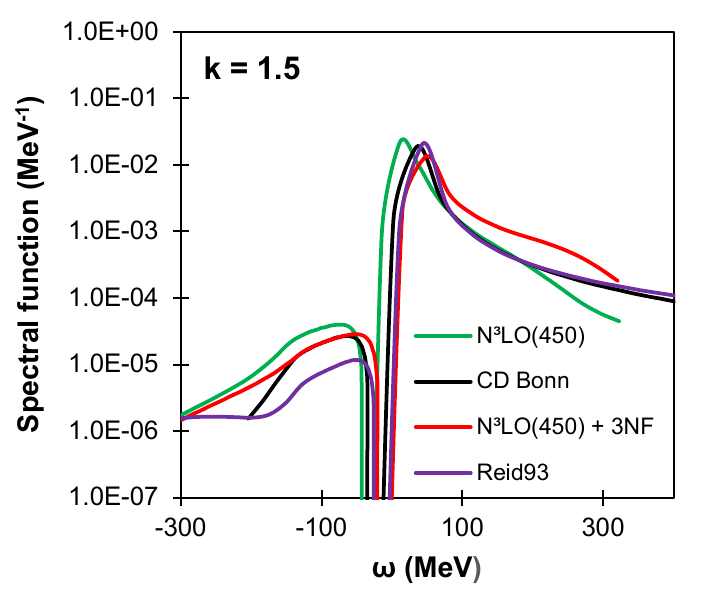}\hspace{0.01in} 
\vspace*{0.05cm}
 \caption{(Color online) Spectral function in nuclear matter with Fermi momentum equal to 1.35 fm$^{-1}$ with different interactions as noted inside the figure. The left and the right panels show the results for $k=0.6 k_F$ and $k=1.5 k_F$, respectively. } 
 \label{sph}
\end{figure*}

We now move to the spectral function, Eq.~(\ref{spctr}), which will lead to the momentum distribution. The high-energy part of the spectral function shows large and characteristic differences between the predictions from chiral interactions and those from CD-Bonn or Reid93, which can be attributed to the softer nature of the chiral forces (cf. Fig.~(\ref{sph})).

\begin{figure*}[!t] 
\centering
\vspace*{1.0cm}
\hspace*{-1.5cm}
\includegraphics[width=6.5cm]{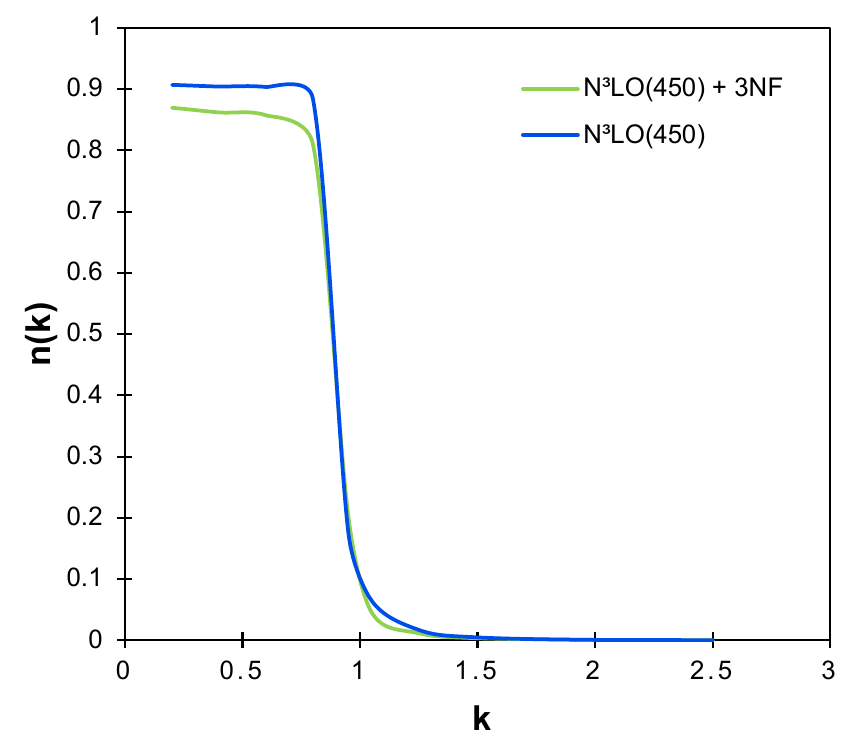}\hspace{0.01in} 
\includegraphics[width=6.7cm]{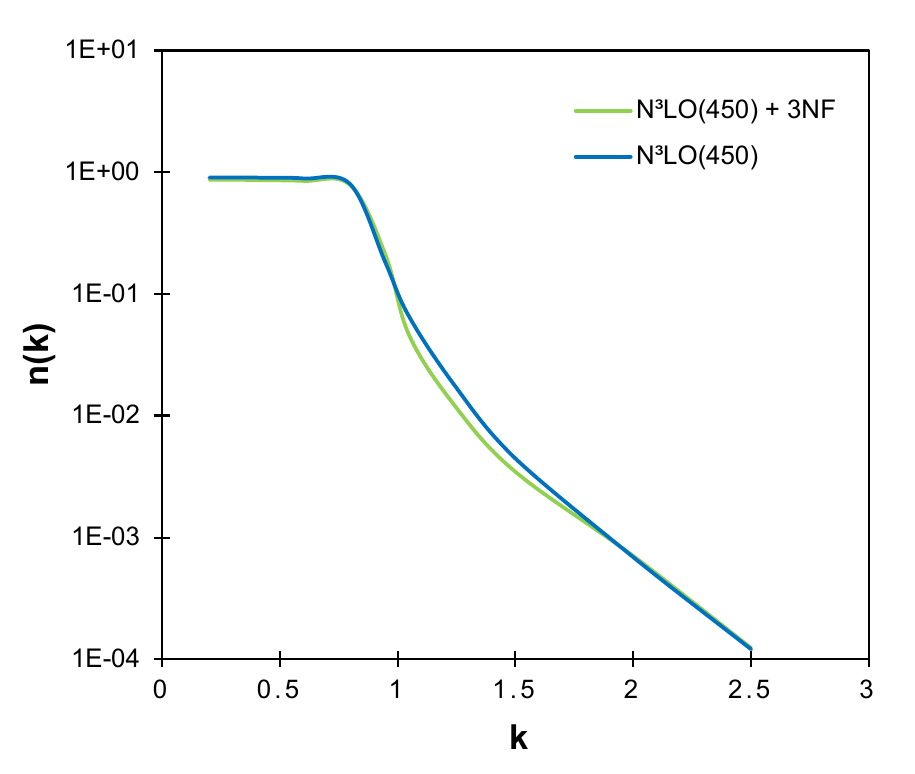}\hspace{0.01in} 
\vspace*{0.05cm}
 \caption{(Color online) Momentum distribution in nuclear matter with Fermi momentum equal to 1.35 fm,$^{-1}$ with and without the inclusion of 3NF. Results are shown on both a linear and a logarithmic scale.} 
 \label{nk}
\end{figure*}

Also, we calculate the momentum density of nucleons in nuclear matter at approximately saturation density, see Fig.~\ref{nk}. In spite of the large sensitivity  to the 3NF we observed in the self-energy, the impact of the 3NF on the momentum distribution is very moderate, probably due to competing effects -- see Figs.~\ref{sig_3nf_04} and \ref{dsig_3nf_04} -- on Eq.~(\ref{spctr}) and consequently on Eq.~(\ref{momdis}).
Note that we use both a linear scale and a logarithmic one to highlight differences in different momentum regions.
The momentum density is slightly reduced by the presence of the chiral 3NF -- a reflection of its repulsive nature.

\begin{figure*}[!t] 
\centering
\vspace*{1.0cm}
\hspace*{-1.5cm}
\includegraphics[width=6.5cm]{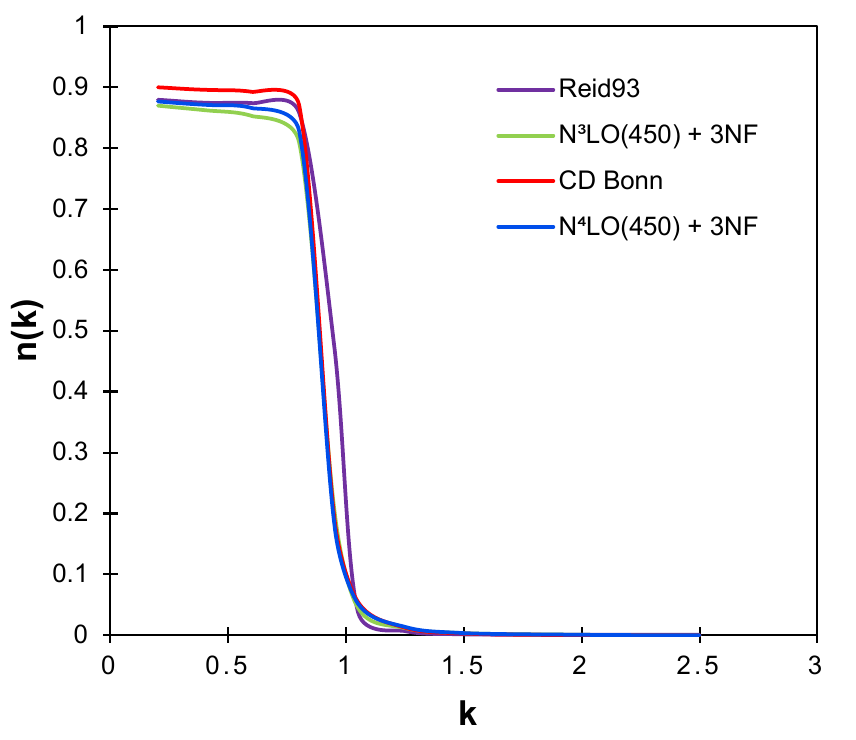}\hspace{0.01in} 
\includegraphics[width=6.8cm]{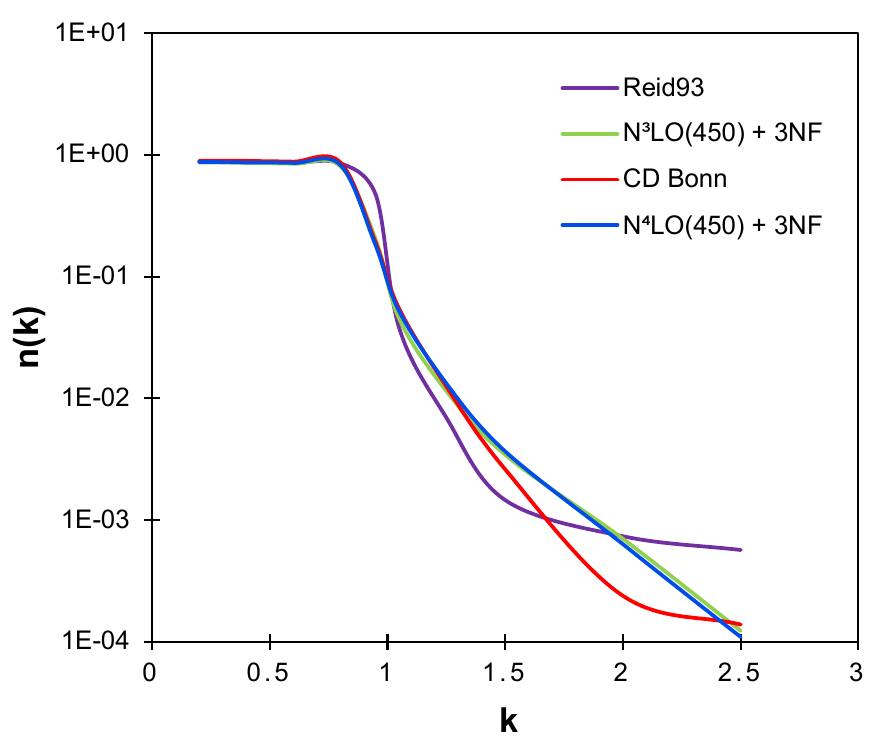}\hspace{0.01in} 
\vspace*{0.05cm}
 \caption{(Color online) Momentum distribution in nuclear matter with Fermi momentum equal to 1.35 fm,$^{-1}$ with and without the inclusion of 3NF, compared with the predictions from CD Bonn and from the Reid93 potential~\cite{Reid93}. Results are shown on both a linear and a logarithmic scale.} 
 \label{nk2}
\end{figure*}

In Fig.~\ref{nk2}, we compare predictions for the momentum distribution at normal density at fourth and fifth orders of the chiral expansion. In both cases, the appropriate 3NF are included as described in sect.~\ref{III}. In addition, we show the predictions obtained with the CD-Bonn potential and the local Reid93 potential. The purpose of this analysis is to estimate the truncation uncertainty at N$^3$LO by comparing with the next  available order. As noted earlier, the predictions at the fourth and fifth order are nearly indistinguishable.

Another measure for the interaction in the nuclear medium is the occurence of quasi-bound two-nucleon states or the formation of pairing correlations. These kind of correlations show up in a very transparent way if one tries to determine the generalized $T$-matrix using the $pphh$ RPA as discussed above. In this framework, the phenomenon of pairing shows up in the form of complex eigenvalues in the solution of the RPA equation. From the eigenvalues and eigenstates of these complex solutions one can easily evaluate the pairing gap function as has been shown in Ref.~\cite{rubts17} and briefly discussed in subsection \ref{sec.pairing}. 

When discussing pairing effects in nuclear physics, it is customary to think of the proton-proton or neutron-neutron pairing observed in even-even nuclei. Indeed, in studies of nuclear matter one can find non-trivial solutions of the BCS equation, which corresponds to complex solution of the $pphh$ RPA equations in the $^1S_0$ partial wave. This pairing in a partial wave with isospin $T=1$ is particularly noticible at low density. Therefore we present results for the gap function in the $^1S_0$ partial wave at approximately one half of the saturation density. As we are mainly interested in comparing results from various interaction models, we have approximated the single-particle spectrum using the parameterization of Eq.~(\ref{spe}) assuming $m^*=m$ in all cases. This leads to larger values of the gap function than using a more realistic spectrum with $m^* < m$. Note that the typically reported gap parameter corresponds to the value of the gap function at $k=k_F$.

\begin{figure*}[!t] 
\centering
\vspace*{1.0cm}
\includegraphics[width=12cm]{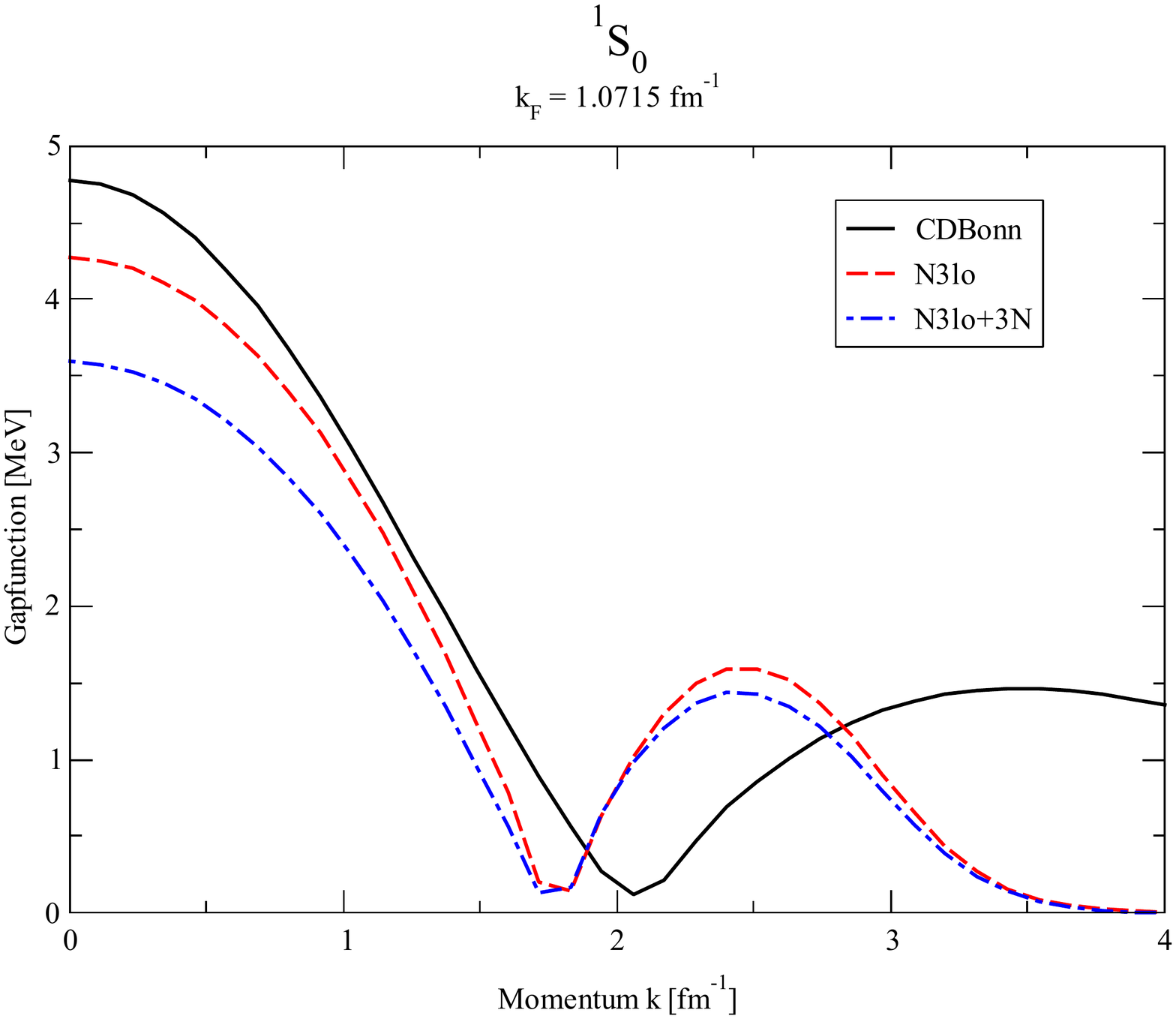}
\vspace*{0.05cm}
 \caption{(Color online) Gap function in the $^1S_0$ partial wave in nuclear matter at half the empirical saturation density using various interaction models.}
 \label{gap1s0}
\end{figure*}

Comparing the results for the various interactions displayed in Fig.~\ref{gap1s0}, one finds that the gap functions determined for the chiral interaction models vanish for value of the momentum $k$ around 4 fm$^{-1}$ and larger, whereas the gap function for the CD-Bonn interaction extends to higher momenta. This is again a consequence of the regulator in the chiral interaction model. At small momenta all three interaction models considered show a similar qualitative behavior, with larger values for the gap function predicted by CD-Bonn. Note that the repulsive 3NF reduces the gap function considerably.

The pairing effect in the $^1S_0$ partial wave is due to the attractive interaction, which is almost sufficient to provide a bound state for two neutrons in the vacuum. The attraction in the $NN$ interaction, however, is even stronger in the $^3S_1-^3D_1$ partial wave, leading to the deuteron. Indeed much larger values for the gap parameter are obtained solving the $pphh$ RPA equations using realistic $NN$ interactions in this channel. The reasons, why these pairing phenomena of proton-neutron pairing are not visible in nuclei have been discussed in Ref.~\cite{MP19}.

\begin{figure*}[!t] 
\centering
\vspace*{1.0cm}
\includegraphics[width=12cm]{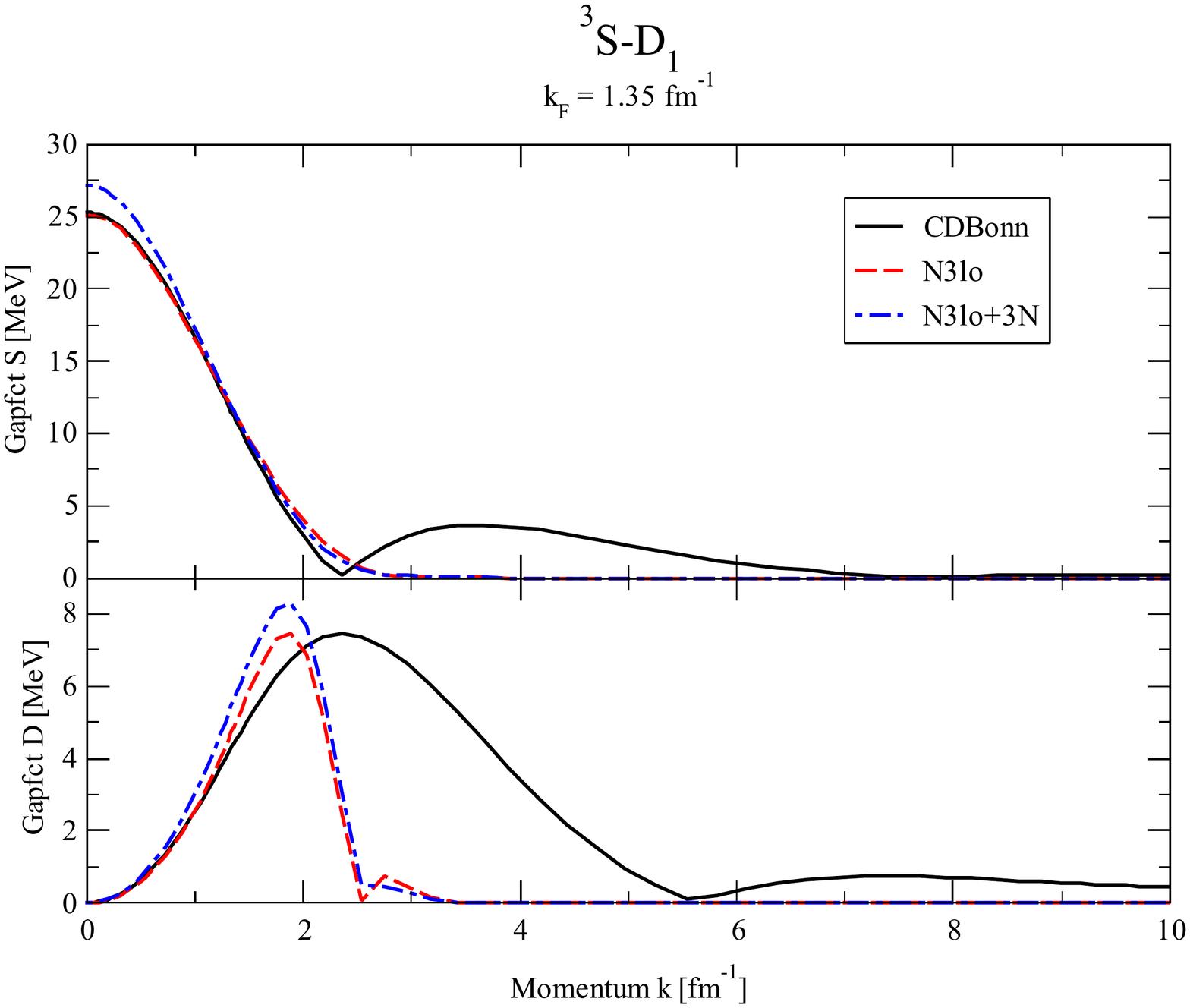}
\vspace*{0.05cm}
 \caption{(Color online) Gap function in the $^3S_1-^3D_1$ partial wave in nuclear matter at  the empirical saturation density using various interaction models. The upper part shows the contribution of the $^3S_1$ part, while the lower part shows results for $^3D_1$.}
 \label{gap3s1}
\end{figure*}

Results for the pairing functions in this partial wave are presented in Fig.~\ref{gap3s1}. Also in this case the gap function derived from the chiral interaction models are different from zero only for momenta below 4 fm$^{-1}$. In this case the 3NF derived for the chiral interaction model slightly enhances the values as compared to the N$^{3}$LO interaction.The conclusions to be taken from this figure are very similar to those discussed for the $^1S_0$ partial wave.

Nontrivial solutions of the gap equation are also observed in the $^3P_2 - ^3F_2$ channel at densities above nuclear matter saturation density. The values for the gap function are smaller than for the partial waves discussed so far. Conclusions from a comparison of the various interacction models are similar to the ones disscussed above.

\vspace*{1.0cm}

\section{Conclusions}                                                                 
\label{Concl} 

While predictions of the bulk properties of nuclear matter are very important to explore and compare the features of different interactions, the study of single-particle properties in the medium provides a more microscopic insight in what ``nucleons do" in nuclei and nuclear matter. This is an exciting -- though not directly observable --  aspect of nuclear physics.

We presented predictions of the nucleon self-energy, spectral function, and momentum distribution  
in nuclear matter with varying energy, momentum, and density. The self-energy obtained from the lowest order of the hole-line expansion, namely the BHF approximation, is complemented with the hole-hole contribution calculated perturbatively. 

We utilized high-quality chiral interactions and compared predictions with those obtained from traditional meson-exchange nucleon-nucleon potentials, both local and non-local. We find strong and characteristic differences especially in the high-energy part of the predictions.

We observe a good convergence pattern at the fourth order of the chiral expansion and find the impact of the chiral 3NF to be weak. It is important to note, though, that only the leading chiral 3NF is included at this time. A systematic study of the impact of 3NFs up to N$^3$LO is in progress.

Our interest in correlations includes both SRC and pairing correlations. 
Thus, we studied the pairing gap in various channels. The strength of correlations and the range of energy or momentum over which they are significant is known to vary with the interaction~\cite{MD05}. We confirm that to be true, especially between chiral EFT-based interactions and the more traditional, harder potentials of the past, at the higher momenta.
We also observed that the repulsive chiral 3NF reduces the gap function considerably.

We would like to close with some comments announced in the Introduction concerning whether
 occupation numbers and momentum distributions can (in principle) be extracted from experiments. The idea is not new--it goes back to the 80's. References~\cite{FS88,Frank93,Tang,CLAS,CLAS2,CLAS3,Pia+,Eg+06,Shneor,Subedi,Baghda,Pia13,Korover,Hen+17,CT+,Atkwim19} provide an overview of the experimental efforts and discussions which have taken place over the past few decades.

The basic idea is that ``hard scattering" on nuclei, such as the exclusive scattering $A(e,e'pN)$ or inclusive quasielastic electron scattering, can probe the distribution of nucleons inside nuclei. Applying the impulse approximation (IA) as well as other approximations, such as neglecting final state interactions (FSI), 
the ratio of cross sections for scattering off two nuclei is shown to reduce to a quantity to be associated with the ``SRC probabilty" in a particular nucleus.
Of course, the approximations were analyzed and found  to be appropriate for the kinematical region selected by the experiment. In this way, absolute SRC probabilities for various nuclei were extracted (see, for instance, 
Ref.~\cite{Eg+06}), and sometimes even extrapolated to nuclear matter~\cite{Pia13}. From the theoretical standpoint, the measured SRC probabilty would be comparable with the integral of the predicted momentum density, $n(k)$, starting at some minimum momentum~\cite{Eg+06}. Naturally, such predictions can vary wildly depending on the off-shell nature of the chosen interaction. 
While these discussions are interesting within a particular framework, one must keep in mind that momentum densities are, in principle, not observable. That is, ambiguities in the measurements are inherent and not due to approximations or experimental limitations. This is very clearly shown in Ref.~\cite{FH02} from the perspective  of EFT. The ``impossibilty of measuring off-shell effects" was discussed earlier by Fearing and Scherer~\cite{FS00}.

\section*{Acknowledgments}
This work was supported in part by 
the U.S. Department of Energy, Office of Science, Office of Basic Energy Sciences, under Award Number DE-FG02-03ER41270 and by the Deutsche Forschungsgemeinschaft (DFG) under contract no. Mu 705/10-2.


\begin{references}     

\bibitem{MP19} H. M{\" u}ther and A. Polls, Phys. Rev. C {\bf 99}, 034315 (2019).


\bibitem{FGM02} T. Frick, Kh. Gad, and H. M{\" u}ther, Phys. Rev. C {\bf 65}, 034321 (2002).

\bibitem{SP86} R. Schiavilla and V.R. Pandharipande, Nucl. Phys. {\bf A449}, 219 (1986).

\bibitem{BCdALS86} O. Benhar, C. Ciofi degli Atti, S. Liuti, and G. Salme`, Phys. Lett. B  {\bf 177}, 135 (1986).

\bibitem{CdAPS91} C. Ciofi degli Atti, E. Pace, and G. Salme`, Phys. Rev. C {\bf 43}, 1155 (1991).


\bibitem{Ciof} C. Ciofi degli Atti and S. Simula, Phys. Rev. C {\bf 53}, 1689 (1996).

\bibitem{MP99}  H. M{\" u}ther and A. Polls, Phys. Rev. C {\bf 61}, 014304 (1999).

\bibitem{MP00}  H. M{\" u}ther and A. Polls, Prog. Part. Nucl. Phys. {\bf 45}, 243 (2000).

\bibitem{A+13} M. Alvioli, C. Ciofi degli Atti, L.P. Kaptari, C.B. Mezzetti, and H. Morita, Phys. Rev. C {\bf 87}, 034603 (2013).

\bibitem{src2015} F. Sammarruca, Phys. Rev. C {\bf 92}, 044003 (2015).      

\bibitem{A+16} M. Alvioli, C. Ciofi degli Atti, and H. Morita, Phys. Rev. C {\bf 94}, 044309 (2016).

\bibitem{MSVM19} L.E. Marcucci, F. Sammarruca, M. Viviani, and R. Machleidt, Phys. Rev. C {\bf 99}, 034003 (2019).

\bibitem{MD05}  H. M{\" u}ther and W.H. Dickoff, Phys. Rev. C {\bf 72}, 054313 (2005).

\bibitem{FS88} L. Frankfurt and M. Strikman, Phys. Rep. {\bf 160}, 235 (1988).

 \bibitem{Frank93} L. L. Frankfurt, M.I. Strikman, D.B. Day, and M. Sargsyan, Phys. Rev. C {\bf 48}, 2451 (1993).

 \bibitem{Tang} A. Tang {\it et al.}, Phys. Rev. Lett. {\bf 90}, 042301 (2003).  

\bibitem{CLAS} K.S. Egiyan {\it et al.}, Phys. Rev. Lett. {\bf 96}, 082501 (2006), and references therein.

\bibitem{CLAS2} K.Sh. Egiyan {\it et al.}, CLAS-NOTE 2005-004, 2005, www1.jlab.org/ul/Physics/Hall-B/clas.

\bibitem{CLAS3} K.Sh. Egiyan {et al.}, Phys. Rev. C {\bf 68}, 014313 (2003). 


\bibitem{Pia+} E. Piasetzky {\it et al.}, Phys. Rev. Lett. {\bf 97}, 162504 (2006).

\bibitem{Eg+06} K.S. Egiyan {\it et al.}, Phys. Rev. Lett. {\bf 96}, 082501 (2006).

\bibitem{Shneor} R. Shneor {\it et al.} (Jefferson Lab Hall A Collaboration), Phys. Rev. Lett. {\bf 99}, 072501 (2007).

\bibitem{Subedi} R. Subedi {\it et al.}, Science {\bf 320}, 1476 (2008).

\bibitem{Baghda} H. Baghdasaryan {\it et al.} (CLAS Collaboration), Phys. Rev. Lett. {\bf 105}, 222501 (2010). 


\bibitem{Pia13} E. Piasetzky, O. Hen, and L.B. Weinstein, AIP Conf. Proc. 1560, 355 (2013). 


\bibitem{Korover} I. Korover {\it et al.} (Jefferson Lab Hall A Collaboration), Phys. Rev. Lett. {\bf 113}, 022501 
(2014). 

\bibitem{Hen+17} Or Hen, G.A. Miller, E. Piasetzky, and L.B. Weinstein, Rev. Mod. Phys. {\bf 89}, 045002 (2017).

\bibitem{CT+} R. Cruz-Torres {\it et al.}, Phys. Lett. B {\bf 797}, 134890 (2019).

\bibitem{Atkwim19} M.C. Atkinson and W.H. Dickhoff, Phys. Lett. B {\bf 798}, 135027 (2019).

\bibitem{EMN17} 
D. R. Entem, R. Machleidt, and Y. Nosyk,
Phys. Rev. C {\bf 96}, 024004 (2017).

\bibitem{Hofe+} M. Hoferichter, J. Ruiz de Elvira, B. Kubis, and U.-G. Meissner, Phys. Rev. Lett. {\bf 115}, 192301 (2015); Phys. Rep. {\bf 625}, 1 (2016). 


 \bibitem{Hop17}
  J.~Hoppe, C.~Drischler, R.~J.~Furnstahl, K.~Hebeler, and A.~Schwenk,
  Phys.\ Rev.\ C {\bf 96}, 054002 (2017).


\bibitem{DHS19} 
  C.~Drischler, K.~Hebeler and A.~Schwenk,
  Phys.\ Rev.\ Lett.\  {\bf 122}, 042501 (2019)

 \bibitem{Epe02} 
E. Epelbaum, A. Nogga, W. Gl\"ockle, H. Kamada, U.-G. Mei\ss ner, and H. Witala,
{Phys. Rev. C} {\bf 66}, 064001 (2002).

\bibitem{holt09} J. W. Holt, N. Kaiser, and W. Weise, Phys. Rev. C {\bf 79}, 054331 (2009).

\bibitem{holt10} J. W. Holt, N. Kaiser, and W. Weise, Phys. Rev. C {\bf 81}, 024002 (2010).

\bibitem{Ber08} V. Bernard, E. Epelbaum, H. Krebs, and Ulf-G. Mei\ss  ner, Phys. Rev. C {\bf 77}, 064004 (2008); {\bf 84}, 054001 (2011).




\bibitem{Dri16} 
  C.~Drischler, A.~Carbone, K.~Hebeler, and A.~Schwenk,
  Phys.\ Rev.\ C {\bf 94}, 054307 (2016).
  
\bibitem{Tew13}
I.\ Tews, T.\ Kr\"{u}ger, K.\ Hebeler, and A.\ Schwenk, Phys.\ Rev.\ Lett.\ {\bf 110}, 032504 (2013).

\bibitem{Heb15a}
  K.~Hebeler, H.~Krebs, E.~Epelbaum, J.~Golak and R.~Skibinski,
  Phys.\ Rev.\ C {\bf 91}, 044001 (2015).

\bibitem{Kais18}  N. Kaiser and V. Niessner, Phys. Rev. C {\bf 98}, 054002 (2018).

\bibitem{Kais19}  N. Kaiser and B. Singh, Phys. Rev. C {\bf 100}, 014002 (2019).     


\bibitem{Dickhoff} W. H. Dickhoff and D. Van Neck, {\em Many-Body Theory
Exposed!} (World Scientific, Singapore, 2005).
\bibitem{ramos1} A. Ramos, A. Polls, and W.H. Dickhoff, Nucl. Phys. A{\bf 503}, 1 (1989).
\bibitem{frick05} T. Frick, H. M{\"u}ther, A. Rios, A. Polls, and A. Ramos, Phys. Rev. C
{\bf 71}, 014313 (2005).
\bibitem{rubts17} O.A. Rubtsova, V.I. Kukulin, V.N. Pomerantsev, and H. M\"uther, Phys. Rev. C{\bf 96}, 034327 (2017).
\bibitem{NM1} H. M{\"u}ther, O.A. Rubtsova, V.I. Kukulin, and V.N.
Pomerantsev, Phys. Rev. C {\bf 94}, (2016).


\bibitem{CD} R. Machleidt, Phys. Rev. C {\bf 63}, 024001 (2001). 

\bibitem{Reid93} V.G.J. Stoks, R.A.M. Klomp, C.P.F. Terheggen, and J.J. de Swart, Phys. Rev. C {\bf 49}, 2950 (1994).   


\bibitem{Machxxx} R. Machleidt, Adv. Nucl. Phys. {\bf 19}, 189 (1989).

\bibitem{eriklowk} E.N.E. van Dalen and H. M\"uther, Phys. Rev. C {\bf 80}, 037303 (2009).


\bibitem{FH02} R.J. Furnstahl and H.-W. Hammer, Phys. Lett. B {\bf 531}, 203 (2002).
\bibitem{FS00} H.W. Fearing and S. Scherer, Phys. Rev. C {\bf 62}, 034003 (2000).
\end{references}
\end{document}